\documentclass[twocolumn]{emulateapj}

\usepackage{amsmath}

\defcitealias{RSR02}{RSR}
\defcitealias{WSR+08}{WSR}

\shortauthors{
Wong at al.}
\shorttitle{
Non-equipartition Effects on Cosmology
}
\slugcomment{Astrophysical Journal, accepted}

\begin{document}
\title{
The Impact of Non-Equipartition on Cosmological Parameter 
Estimation from Sunyaev--Zel'dovich Surveys
}

\author{Ka-Wah Wong, 
Craig L. Sarazin, 
and 
Daniel R. Wik
}

\affil{Department of Astronomy, University of Virginia,
P.O. Box 400325, Charlottesville, VA 22904-4325, USA
}

\email{kwwong@virginia.edu, sarazin@virginia.edu, drw2x@virginia.edu}

\begin{abstract}
The collisionless accretion shock at the outer boundary of a galaxy 
cluster should primarily heat the ions instead of electrons since they 
carry most of the kinetic energy of the infalling gas.  Near the 
accretion shock, the density of the intracluster medium is very 
low and the Coulomb collisional timescale is longer than the accretion 
timescale.  Electrons and ions may not achieve equipartition in these 
regions. 
Numerical simulations have shown that the Sunyaev--Zel'dovich 
observables
(e.g., the integrated Comptonization parameter $Y$) for relaxed clusters
can be biased by a few percent.
The $Y$ versus mass
relation can be biased if non-equipartition effects are not properly taken 
into account.  Using a set of hydrodynamical simulations we have 
developed, we have calculated three potential systematic biases in the 
$Y$ versus mass relations introduced by non-equipartition effects during 
the cross-calibration or self-calibration when using the galaxy cluster 
abundance technique to constrain cosmological parameters.  We then use a 
semi-analytic technique to estimate the non-equipartition effects on the 
distribution functions of $Y$ ($Y$ functions) determined from the 
extended Press--Schechter theory.
Depending on the calibration method, we 
find that non-equipartition effects can induce systematic biases on the $Y$ 
functions, and the values of the cosmological parameters $\Omega_M, 
\sigma_8$, and the dark energy equation of state parameter $w$ can be 
biased by a few percent.  In particular, non-equipartition effects can 
introduce an apparent evolution in $w$ of a few percent in all of the 
systematic cases we considered.  
Techniques are suggested to take into account the 
non-equipartition effect empirically when using the cluster abundance 
technique to study precision cosmology. 
We conclude that systematic 
uncertainties in the $Y$ versus mass relation of even a few percent can 
introduce a comparable level of biases in cosmological parameter 
measurements. 
\end{abstract}


\section{Introduction}
\label{sec:intro}

Observational and theoretical studies have shown that galaxy clusters can 
be used as cosmological probes.  In particular, the 
evolution of the galaxy cluster abundance, or the mass function, is 
sensitive to cosmological parameters including the average matter density 
$\Omega_M$, 
the normalization of the power spectrum of the initial density 
fluctuations $\sigma_8$, and the dark energy equation of state parameter 
$w$ \citep{MAE+08,Vik+09}.
Except for gravitational lensing which 
is difficult to do for large sample of galaxy clusters, the masses of 
galaxy clusters cannot be directly measured, and hence the mass function 
cannot be measured easily.
Very often, the masses are estimated using 
mass proxies such as the Sunyaev--Zel'dovich (SZ) temperature distortion, 
the X-ray flux, or galaxy dynamics. 
The mass--observable relations have to be calibrated 
empirically or semi-empirically with numerical simulations 
\citep[e.g.,][]{Vik+09}.  Hence, measuring cosmological parameters using
the galaxy cluster abundance requires a full understanding 
of mass--observable relations.  Even if the mass function and the 
mass--observable relation can be fitted simultaneously 
\citep[``self-calibration'';][]{LSW02,Hu03,MM03}, 
the correct form of the mass--observable relation is needed.

A recent X-ray survey has shown that even a sample of only 85 X-ray 
clusters is sufficient to provide very tight constrains on some 
cosmological parameters.
For example, $\sigma_8$ can be measured down to 1\% level 
in statistical 
uncertainty using the cluster abundance technique alone by assuming a 
flat universe with fixed dark energy equation of state parameter and a 
prior on the Hubble constant \citep{Vik+09}.
However, the statistical 
uncertainties on some other cosmological parameters (e.g., $\Omega_M$) 
are still slightly larger than 10\%.
Ongoing and future SZ surveys will detect thousands of clusters 
\citep[e.g.,][]{Bir99,CHR02,BCM+08}, and this will significantly 
improve the constraints on cosmological parameters.  
Therefore, it is important 
to control the systematic uncertainties of galaxy cluster physics at even 
a percentage level.

Because of the very long Coulomb collisional timescale in the 
low-density outer 
regions of galaxy clusters, it has been pointed out that electron and 
ions there can be in non-equipartition \citep{FL97,EF98}.  Numerical 
simulations have shown that the SZ observables, e.g., integrated 
Comptonization parameter ($Y$), for relaxed clusters can be biased by a 
few percent \citep{WS09}, 
and can potentially be biased up to $\sim 10\%$ in 
major merging clusters \citep{RN09}.  Specifically, the non-equipartition 
effect reduces the electron pressure compared to equipartition models, 
and hence the integrated Comptonization parameter of the 
non-equipartition model is smaller than that of the equipartition model.  
Although the uncertainties are still large, recent X-ray observations 
suggest that the electron pressure in cluster outer regions may be lower 
than that predicted by numerical simulations assuming equipartition 
\citep{Bas+09,GFS+09,Hos+10}.  Recent observations of the secondary 
cosmic microwave background anisotropies with the South Pole Telescope 
(SPT) and the {\it WMAP} 7 year data also suggest that the electron 
pressure 
is smaller than the value predicted by hydrodynamic simulations 
\citep{Lue+09,Kom+10}.  These observational signatures are consistent 
with electrons and ions in non-equipartition, although it is also 
possible that the hydrodynamic simulations may simply overestimate the 
gas pressure.  
Another possibility is that heat conduction outside the clusters may be 
reducing the gas pressure \citep{Loe02}.

In our previous paper, we have shown that the non-equipartition effect 
can 
introduce biases in the integrated SZ effect, and the biases depend on 
cluster mass and evolve with redshift in the $\Lambda$CDM cosmology.  The 
non-equipartition model was discussed in detail in \citet{WS09}.  In this 
paper, we study the impact of non-equipartition effect on precision 
cosmology studies using the non-equipartition models we have developed. 
We consider only the non-equipartition effects associated with an accretion
shock at the outer edge of a cluster.
No significant collisionless electron heating at the accretion shocks is assumed,
and the intergalactic gas outside the collisionless accretion shock is taken to be cold.
These assumptions maximize the non-equipartition effects of the accretion shock.
On the other hand, cluster mergers are not considered in our work.  
Mergers may increase the non-equipartition effect by a few percent, and 
hence our calculations of the biases in cosmological parameter estimation 
may still underestimate these effects.  However, the non-equipartition 
effect induced by mergers lasts for only 0.5--1~Gyr \citep{RN09} which 
is comparable to the timescale that mergers can temporarily enhance the 
integrated SZ effect by boosting the overall temperature 
\citep[][hereafter WSR]{WSR+08}. 
Such transient phenomenon will mainly
introduce scatter in the $Y$ versus mass relation, and the effect on 
cosmological parameter estimation is small in general.  The 
non-equipartition effect induced by mergers may partially cancel out the 
merger boost in SZ effect, and hence the merger effect on cosmological 
parameter estimation may even be smaller.  On the other hand, the 
non-equipartition effects in the accretion shock regions can 
systematically bias the $Y$ versus mass relation for all clusters, as 
long as clusters are continuously accreting materials from the 
surrounding which is believed to be generally true.  
Moreover, systematic uncertainties in precision cosmology using galaxy 
clusters can be minimized by restricting the sample of clusters to the 
highest degree of dynamical relaxation, and hence considering the 
systematic effects on relaxed clusters alone is particularly important. 
We follow \citet[][hereafter RSR]{RSR02} 
and \citetalias{WSR+08} closely to quantify the biases 
in cosmological parameter estimation using semi-analytical techniques.  
Specifically, we study the non-equipartition 
effect on the $Y$ versus mass relation (Section~\ref{sec:YM}).  Such a 
biased $Y$ versus mass relation will affect the number of clusters with 
$Y$ observed in SZ surveys, i.e., the $Y$ function 
(Section~\ref{sec:SZsurveys}).  In this work, the $Y$ function is 
calculated using the extended Press--Schechter theory \citep{PS74}.  We 
consider 
three cases which may potentially introduce biases in the $Y$ versus mass 
relations if the non-equipartition effect is not properly taken into 
account during the cross-calibration or self-calibration processes when 
using the galaxy cluster abundance technique (Section~\ref{sec:sys}).  We 
quantify and discuss the impact on cosmological parameter estimation from 
SZ surveys by fitting the mass function with the biased $Y$ versus mass 
relations (Section~\ref{sec:cosmofit}).  Section~\ref{sec:conc} gives the 
discussion and conclusions.  Throughout the paper, we assume the Hubble 
constant $H_0=71.9 \, h_{71.9}$~km~s$^{-1}$~Mpc$^{-1}$ with $h_{71.9}=1$.

\section{SZ Versus Mass Correlation}
\label{sec:YM}
The SZ effect can be characterized as the Comptonization parameter, $y$, 
which is given by
\begin{equation}
\label{eq:y-para}
y = \frac{k_B \sigma_{\rm T}}{m_e c^2} \int n_e T_e dl \propto \int P_e 
dl
\, ,
\end{equation}
where $\sigma_{\rm T}$ is the Thomson scattering cross section, 
$n_e$ is the electron number density, 
$T_e$ is the electron temperature, 
$P_e=n_e k_B T_e$ is the electron pressure,
and $l$ is the distance along the line of sight.
The integrated Comptonization parameter, $Y$, is defined as the 
integral of the Comptonization parameter in equation~(\ref{eq:y-para}) 
over the area of the cluster on the sky
\begin{equation}
\label{eq:bigY}
 Y = d_A^2 \int y d\Omega = \int y dA \, ,  
\end{equation}  
where $d_A$ is the angular diameter distance to the cluster, $\Omega$ is 
the solid angle of the cluster on the sky, and $A$ is the projected 
surface area.  In this paper, $Y$ is integrated over the projected 
surface area of the cluster out to the shock radius.  This quantity is 
useful for spatially unresolved clusters with SZ observations where the 
beam area covers the whole cluster.

\begin{figure}
\includegraphics[angle=270,width=9.0cm]{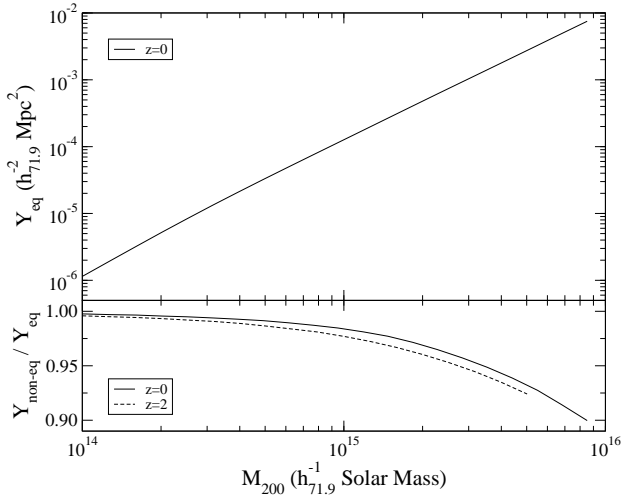}
\caption{Top panel: Integrated Comptonization parameter $Y$ versus mass 
at $z=0$ for the equipartition model used in this paper.  Bottom 
panel: Integrated $Y$ bias, $(Y_{\rm non{\textrm -}eq}/Y_{\rm eq})_{\rm WS}$, 
versus mass at $z=0$ and $2$.
}
\label{fig:YvsM}
\end{figure}

It has been shown that the integrated Comptonization parameter displays a 
tight correlation with cluster mass \citep{RS06}.  Such a tight 
correction is useful for precision cosmology, and hence a correct 
understanding of the integrated Comptonization parameter is important.  A 
detailed discussion of the use of SZ surveys to study cosmology can be 
found in \citet{CHR02}.  In this paper, we assume the SZ effect versus 
mass 
relation for the equipartition model to be the same as the equilibrium 
$Y$--$M$ relation used in \citetalias{WSR+08}.  Specifically, the 
equipartition SZ effect versus mass relation we assume is of the form
\begin{equation}
\label{eq:YeqvsM}
Y_{\rm eq}=N \, x^{\alpha} p[x] \,\, h_{71.9}^{-2}\, {\rm Mpc}^2
\, ,
\end{equation}  
where $x= M_{200}/(h_{71.9}^{-1} 10^{15}  M_{\odot})$, $N$ is the 
normalization constant, $\alpha$ is the power-law index, and $p[x]$ is a 
13 degree polynomial in $x$.  Equation~(\ref{eq:YeqvsM}) is fitted to the 
numerical solutions for the equilibrium $Y$--$M$ relation in 
\citetalias{WSR+08}.  The 
integrated SZ bias introduced by the 
non-equipartition effect, $Y_{\rm non{\textrm -}eq}/Y_{\rm eq}$ versus $M$ at 
different redshifts, is taken from \citet{WS09}.  The non-equipartition 
SZ 
effect versus mass relation we used in this paper is hence given by
\begin{equation}
\label{eq:YnoneqvsM}
Y_{\rm non{\textrm -}eq}= Y_{\rm eq, WSR} \times \left( \frac{Y_{\rm 
non{\textrm -}eq}}{Y_{\rm eq}} \right)_{\rm WS}
\, ,
\end{equation} 
where the subscripts ``WSR'' and ``WS'' here indicate that the terms are 
taken from different models in \citetalias{WSR+08} and \citet{WS09}, 
respectively.
In this work, since we are interested in the relative effects on SZ 
surveys and cosmological parameter estimation introduced by the
non-equipartition effects instead of the precise $Y$--$M$ relation which 
depends on details of numerical simulations, in principle, we can 
take any equilibrium $Y$--$M$ relation from simulations and apply our 
non-equipartition bias to the equilibrium $Y$--$M$ relation. 
The reason for using the equilibrium $Y$--$M$ relation in 
\citetalias{WSR+08} 
to model the non-equipartition $Y$--$M$ relation instead of the $Y$--$M$ 
relation in \citet{WS09} is that the former relation takes into 
account the dependence of $Y$ on gas fraction $f_{\rm gas}$, where 
$f_{\rm gas}\propto M^{1/3}$ for $M_{200}\gtrsim 10^{14} M_{\odot}$.  The 
numerical solutions in \citet{WS09} assume a constant $f_{\rm gas}$, and 
a self-similar argument shows that $Y \propto M^{5/3} f_{\rm gas} 
\propto 
M^{5/3}$.  The equilibrium $Y$--$M$ relation in \citetalias{WSR+08} has a 
power-law index close to 2.  On the other hand, the non-equipartition 
effects on the integrated $Y$ depend weakly on $f_{\rm gas}$ and hence we 
can assume the $Y$ bias of the constant $f_{\rm gas}$ models in 
\citet{WS09} can be applied to the varying $f_{\rm gas}$ model in 
\citetalias{WSR+08}.  Another advantage of using the equilibrium 
$Y$--$M$ 
relation in \citetalias{WSR+08} is that we can compare the effect of 
non-equipartition on cosmological parameters estimations to the merger 
effects calculated in \citetalias{WSR+08}.  The equipartition $Y$--$M$ 
relation and the integrated SZ bias introduced by the non-equipartition 
effect we used in this paper are plotted in Figure~\ref{fig:YvsM}.
Clusters with higher masses are hotter, and hence, the equipartition 
timescales are longer.  Thus, the non-equipartition effects are stronger 
in more massive clusters \citep{FL97, WS09}.   
For our non-equipartition model in the $\Lambda$CDM universe, the 
integrated SZ bias decreases as redshift decreases.  This is probably due 
to the decreasing rate of accretion onto clusters in the $\Lambda$CDM 
universe during the cosmological acceleration, which results in a 
longer time for electron--ion equilibration \citep{WS09}.

In order to quantify the effect of non-equipartition on the SZ versus 
mass relation, we fit a power-law function to the integrated SZ bias of 
the form
\begin{equation}
\label{eq:YbiasvsM}
(Y_{\rm non{\textrm -}eq}/Y_{\rm eq})_{\rm WS} = \Delta N \left( 
\frac{M_{200}}{10^{15} M_{\odot}} \right)^{\Delta\alpha}
\end{equation}
to all the clusters with $M_{200}$ between $10^{14}$ and $4 \times 
10^{15} M_{\odot}$.  The range is consistent with the cluster mass range 
used to study the mass function in X-ray observations \citep{Vik+09}.  We 
also fit over the wider range of $10^{13}$ to $10^{16} M_{\odot}$ for 
comparison.  The fitted coefficients $\Delta N$ and $\Delta\alpha$ 
correspond to the biases in the fitted SZ versus mass relation if the 
non-equipartition effect is not taken into account.  The clusters in our 
sample are distributed roughly uniformly in the logarithm of the mass.  
The fitted results are listed in Table~\ref{tab:YMfit}.

The effect of non-equipartition is to decrease $Y$ for the high-mass 
clusters, and hence the effect on the SZ versus mass relation is to lower 
the power-law index by 0.01 to 0.016, which corresponds to a decrease of 
0.5\%--0.8\% for the SZ versus mass relation with power-law index 
$\alpha = 2$.  This is comparable to the 0.8\% uncertainty of the 
power-law index derived from simulations combined with observations 
within $M_{500}$  \citep{APP+09}.  The normalization 
is biased to compensate for the change in power-law index, and the bias 
is 
about 3\% for the scaled mass of $10^{15} M_{\odot}$.  For SZ 
surveys which measure $Y$ out to the shock radius, e.g., observations 
with 
a spatial resolution poorer than typical shock radii, we have shown that 
the non-equipartition effect can introduce a small deviation in the 
measured $Y$--$M$ relation.  The deviations are small, but future SZ 
surveys with sufficient statistics may be able to detect such signatures.  
On 
the other hand, if clusters are spatially resolved and $Y$ are measured 
within $R_{200}$, we have shown that non-equipartition effect is smaller 
than 1\% for all clusters, and hence the bias in the  $Y$--$M$ relation 
is negligible \citep{WS09}.  A bias in measurement out to the shock 
radii but not within $\sim R_{200}$ will indicate that cluster outer 
regions may be in non-equipartition.

\section{Effects of Non-Equipartition on SZ Surveys}
\label{sec:SZsurveys}
The number of galaxy clusters expected to be found per unit comoving 
volume depends sensitively on cosmology.  The quantity which is 
convenient to describe the cluster number density is the mass function, 
$n(M,z)$, where $n(M,z)dM$ gives the number of clusters per unit comoving 
volume with masses in the range $M\rightarrow M+dM$, and $z$ is the 
redshift.  While the exact form of the mass function can be found most 
accurately from cosmological simulations \citep{Spr+05}, a semi-analytic 
form of the mass function given by the extended Press--Schechter 
theory \citep{PS74, BCE+91, LC93} provides a more convenient way to 
understand the dependence of the mass function on cosmological 
parameters, especially when we are interested in the relative effect 
instead of the precise values of the mass function itself.  Although the 
Press--Schechter theory cannot reproduce the mass function found in 
cosmological simulations at very high redshifts and low cluster masses 
\citep{ST99, LHH+07}, it is more than sufficient over the redshifts 
($z=0$ to $2$) and masses ($M=10^{14}$ to $10^{16} 
M_{\odot}$) of interest here.

The mass function given by the extended Press--Schechter theory can be 
written as \citep{PS74}
\begin{widetext}
\begin{equation} 
\label{eq:nmps}
    n_{\rm PS}(M,z)dM = \sqrt{\frac{2}{\pi}}
    \frac{\bar \rho}{M} \frac{\delta_c(z)}
    {\sigma^2 (M)} \Bigg| \frac{d\sigma(M)}{dM} \Bigg|
    \mathrm{exp} \Biggl[ - \frac{\delta_c^2(z)}
    {2\sigma^2(M)} \Biggr] dM \, ,
\end{equation}
\end{widetext}
where ${\bar \rho}$ is the current mean of the total mass density of the 
universe, $\sigma(M)$ is the current rms density fluctuation within a 
sphere of mean mass $M$, and $\delta_c(z)$ is the critical linear 
overdensity required for a region to collapse at redshift $z$.  Unless 
otherwise specified, the parameters used in this paper are the same as 
those in \citetalias{RSR02} and \citetalias{WSR+08}.  

Once the $Y$ versus mass relation is known, the distribution function of 
$Y$ is given by the $Y$ function,
\begin{equation}
\label{eq:nYps}
n_{\rm PS}(Y,z)= n_{\rm PS}(M,z) \frac{dM}{dY} \, , 
\end{equation}
where $n_{\rm PS}(Y,z)dY$ gives the number of clusters per unit comoving 
volume at redshift $z$ which have the integrated SZ parameters in the 
range $Y \rightarrow Y + dY$.  The $Y$ function for the non-equipartition 
models and the equipartition models can be related by
\begin{equation}
\label{eq:nYnoneq}
n_{\rm PS,non{\textrm -}eq}(Y_{\rm non{\textrm -}eq},z)= n_{\rm PS,eq}(Y_{\rm eq},z) 
\frac{dY_{\rm eq}}{dY_{\rm non{\textrm -}eq}} \, , 
\end{equation}
where the subscripts ``eq'' and ``non-eq'' denote the equipartition and 
the 
non-equipartition models, respectively.

\begin{figure}
\includegraphics[angle=270,width=9.0cm]{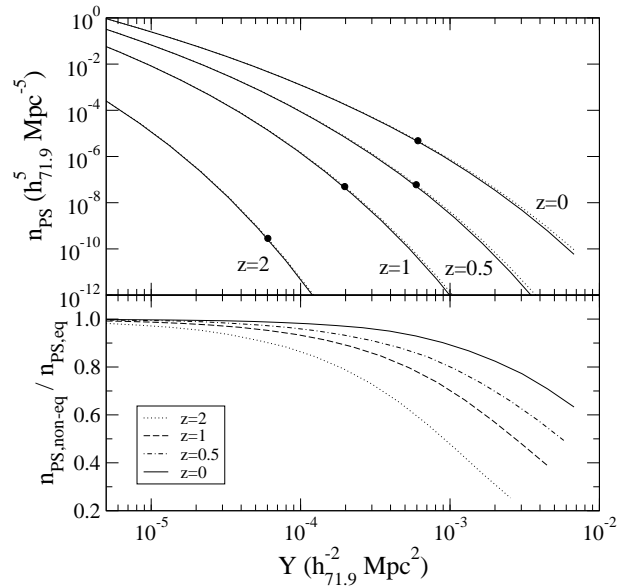}
\caption{Top panel: Integrated $Y$ functions of the non-equipartition 
(solid lines) and the equipartition (dotted lines) models in the standard 
$\Lambda$CDM universe.  Solid dots indicate 
the values of $Y_0$ such that
$\int_{Y_0}^{\infty} \int_{z_l}^{z_u} n_{\rm PS}(Y,z) \, dY \, ( dV / dz ) \, dz = 1$
for the equipartition model.  Bottom panel: Ratios between the 
non-equipartition and  equipartition $Y$ functions.
}
\label{fig:YF}
\end{figure}

Figure~\ref{fig:YF} shows the $Y$ functions for the equipartition and the 
non-equipartition models and also their ratios at different redshifts for 
the standard $\Lambda$CDM cosmology.  The theoretical $Y$ functions can 
be biased strongly for large $Y$ and high redshift clusters.  In 
practice, whether the bias can affect the observed $Y$ functions depends 
on the number of clusters that can be observed, and this depends on 
cosmology. 
The maximum number of clusters that can be observed with $Y$ values in 
the range $Y = Y 
\rightarrow Y + dY$ and with redshifts in the range $z=z \rightarrow z+dz$ is 
$n_{\rm PS}(Y,z)\, dY dV$, where $dV$ is the comoving volume of the 
universe between redshifts $z$ and $z + dz$.
For each of the four redshifts $z = 0$, 0.5, 1.0, and 2.0, we selected a 
redshift interval $z_l$ to $z_u$ as given in the second and third 
columns of Table~\ref{tab:Ylimits}.
Then, we defined values of $Y_0$ and $n_{\rm PS,0}$ such that 
$\int_{Y_0}^{\infty} \int_{z_l}^{z_u} n_{\rm PS}(Y,z) \, dY \, ( dV / dz 
) \, dz = 1$
and $n_{\rm PS,0}(z)=n_{\rm PS}(Y_0,z)$ for the 
equipartition models.
These values are listed 
in Table~\ref{tab:Ylimits} and are plotted in 
Figure~\ref{fig:YF} as solid dots.  
For example, 
between $z = 0.25 \rightarrow 0.75$ for the assumed cosmology,
the expected number of clusters with $Y \gtrsim 
6\times 10^{-4}\, h^{-2}_{71.9}$~Mpc$^2$ 
($M \gtrsim 2\times 10^{15}\, h^{-1}_{71.9} M_{\odot}$)
is one.  On the other 
hand, the expected number of clusters with $Y = (1 \rightarrow 2) \times 
10^{-4}\, h^{-2}_{71.9}$~Mpc$^2$ 
[$M = (0.9 \rightarrow 1.3) \times 10^{15}\, h^{-1}_{71.9} M_{\odot}$] 
is about 200 within the same redshifts interval, and the bias in $n_{\rm 
PS}$ is about 5\%. 

\section{Effects of Non-Equipartition on Cosmological Parameter 
Estimation from SZ Surveys}
\label{sec:cosmo}
In this section, we follow a procedure similar to that outlined in 
\citetalias{RSR02} and \citetalias{WSR+08} to address the impact of 
non-equipartition effects on 
cosmological parameter estimation from SZ surveys.  Readers who are 
interested in the technical details should refer to 
\citetalias{RSR02} and \citetalias{WSR+08}.  We outline the fitting 
procedure used in this work and 
address the difference between the previous works below.  

As discussed in Section~\ref{sec:SZsurveys}, the mass function of galaxy 
clusters is sensitive to cosmology, and hence measuring the galaxy 
cluster abundance at different redshifts can provide constraints to 
cosmological parameters.  In particular, the properties of the dark 
energy can potentially be determined \citep{HMH01}.  However, the masses 
of galaxy clusters cannot be directly determined, and a mass proxy 
must be observed to determine the mass through the mass--observable 
relation.  Examples of mass proxies are the SZ temperature 
distortion, the X-ray flux, and the weak leasing shear.  If the 
mass--observable relations can be calibrated, this can provide very tight 
constraints on cosmological parameters \citep{MAE+08,Vik+09}.  The 
mass--observable relations can be calibrated by numerical simulations 
and/or cross-calibrations with some other observables, and these are 
subjected to systematic uncertainties due to cluster physics and/or 
observational constraints.  On the other hand, the mass--observable 
relations can be simultaneously fitted with the mass function of galaxy 
clusters, and this is called ``self-calibration'' 
\citep{LSW02,Hu03,MM03}.  The sensitivity in constraining cosmological 
parameters and the mass--observable relations by ``self-calibration'' 
depends on both the form of the mass--observable relations and the mass 
function of galaxy clusters.  In this work, we are mainly interested in 
the biases introduced by the non-equipartition effect, and the 
effects on the mass--observable relations.  Hence, we assume there is no 
other systematic uncertainties in the mass function in any given 
cosmology.

\subsection{Systematic Uncertainties Introduced by Non-Equipartition}
\label{sec:sys}

To address the impact of non-equipartition effects on cosmological 
parameter estimation from SZ surveys, we quantify the impact as biases 
in the cosmological parameter estimates if the calibration of the 
$Y$ versus mass relation does not include the non-equipartition effect.
We generate the integrated 
$Y$ function with the non-equipartition effects included under an 
assumed cosmology, and call the generated $Y$ function the 
non-equipartition $Y$ function.  
We then fit the non-equipartition $Y$ function
with the incorrectly calibrated $Y$ versus mass relation.  We consider 
three different cases 
for the incorrectly calibrated $Y$ versus mass relation.  For the first 
case ({\it Case 1}), we assume the $Y$ versus mass relation is calibrated 
with incorrect numerical simulations that assume equipartition, and 
this incorrectly calibrated $Y$ versus mass relation is used to fit 
the mass function.  An example of this systematic bias might occur if 
the $Y$ versus mass relation was extrapolated to the shock radius using 
observations within a smaller radius together with numerical simulations 
assuming equipartition.  In this case, the integrated SZ bias is simply 
given by
\begin{equation}
\label{eq:b1}
b_1 = \left(\frac{Y_{\rm non{\textrm -}eq}}{Y_{\rm eq}}\right)_{\rm WS} \,,
\end{equation}
where the right-hand side is the same term in 
equation~(\ref{eq:YnoneqvsM}).  

For the second case ({\it Case 2}), we 
assume the $Y$ versus mass relation is self-calibrated by fitting the $Y$ 
versus mass relation and the mass function simultaneously, but with an 
incorrect functional form in the $Y$ versus mass relation.  We assume the 
incorrect functional form of the $Y$ versus mass relation to be a 
power law in mass.  In this case, we assume the integrated SZ bias is 
given by
\begin{equation}
\label{eq:b2}
b_2 = \left(\frac{Y_{\rm non{\textrm -}eq}}{Y_{\rm eq}}\right)_{\rm WS} \left / 
\left(\frac{Y_{\rm non{\textrm -}eq}}{Y_{\rm eq}}\right)\right._{\rm plfit} \,,
\end{equation}
where the term with the subscript ``plfit'' is the best-fit power-law 
relation given in equation~(\ref{eq:YbiasvsM}).  

For the third case ({\it 
Case 3}), we assume the $Y$ versus mass relation is calibrated correctly 
at $z=0$, but the $Y$ versus mass relation at higher redshifts is 
incorrectly calibrated by extrapolating the calibration from that at 
$z=0$.  In this case, we assume  the integrated SZ bias is given by
\begin{equation}
\label{eq:b3}
b_3 = \left(\frac{Y_{\rm non{\textrm -}eq}}{Y_{\rm eq}}\right)_{{\rm WS}, z} \left/ 
\left(\frac{Y_{\rm non{\textrm -}eq}}{Y_{\rm eq}}\right)\right._{{\rm WS}, z=0} \,.
\end{equation}
Studying the impact on cosmological parameter estimation for all of these 
cases is equivalently to fitting the non-equipartition SZ luminosity 
function by the equipartition SZ luminosity function in 
equation~(\ref{eq:nYnoneq}), but replacing the $dY_{\rm eq}/dY_{\rm 
non{\textrm -}eq}$ by $b_1, b_2$, and $b_3$ in 
equations~(\ref{eq:b1})--(\ref{eq:b3}).

\subsection{Fitting Procedures and Results}
\label{sec:cosmofit}

For each case of the systematic bias we studied, we generate the $Y$ 
function with the non-equipartition bias included as given in 
equation~(\ref{eq:nYnoneq}) by assuming the standard $\Lambda$CDM 
cosmological model with $\Omega_M=0.258$, $\sigma_8 = 0.796$, and a 
constant dark energy equation of state parameter $w=-1$.  The $Y$ 
function generated can be directly calculated from the analytic 
Press--Schechter mass function in equation~(\ref{eq:nmps}) and the biased 
$Y$ versus mass relations.  This is simpler than those 
in \citetalias{RSR02} or \citetalias{WSR+08} where merger trees were 
needed to generate the mass 
function in order to follow the merger history.  In 
our work, all the $Y$ versus mass relations are biased regardless of the 
merger history.  We then fit the generated $Y$ function with an 
equipartition model and find the best-fit values of the cosmological 
parameters.  Clusters of galaxies can be used to constrain 
the dark energy equation of state parameter, $w$.  Even the evolution of 
$w$ can potentially be constrained.  In this work, we study the 
constraint on $w(z)$ using the form $w=w_0+w_1 z/(1+z)^2$; the detailed 
explanation of this choice can be found in \citetalias{WSR+08}.  We 
consider three cases when fitting the cosmological parameters: i) fitting 
the $\Omega_M$ and $\sigma_8$ but fixing $w(z)=-1$; ii) fitting the 
$\Omega_M$, $\sigma_8$, and assuming $w(z)=w_0$, where $w_0$ is a 
constant to be fitted; and iii) fitting the $\Omega_M$, $\sigma_8$, and 
$w=w_0+w_1 z/(1+z)^2$, where $w_0$ and $w_1$ are constants to be fitted.  
The $Y$ functions are simultaneously fitted at four different redshifts 
($z=0, 0.5, 1.0$, and $2.0$) to break the degeneracy in the fitted 
cosmological parameters.  We choose only to fit $Y$ between $5\times 
10^{-6} \, h_{71.9}^{-2}$~Mpc$^2$ and $5\times 10^{-3} \, 
h_{71.9}^{-2}$~Mpc$^2$.  
The lower limit is chosen because clusters are likely to be confused in 
the SZ surveys for $M\lesssim 10^{14} \, h_{71.9}^{-1} M_{\odot}$ 
\citep{HMB07}.  Increasing the lower limit will make the biases in 
cosmological parameters larger since the non-equipartition effect 
increases 
with cluster mass.  The upper limit corresponds roughly to the most 
massive cluster that can be formed in the $\Lambda$CDM universe.  The 
limits are also consistent with SZ surveys being done or planned 
\citep[e.g.,][]{MBD05}. 
We also limit the fits to values of $Y < Y_0$ and $z$ such that the 
maximum 
number of observable clusters
$\int_{Y_0}^{\infty} \int_{z_l}^{z_u} n_{\rm PS}(Y,z) \, dY \, ( dV / dz ) \, dz  \ga 1$
for most  redshift bins (Table~\ref{tab:Ylimits}).

\begin{figure*}
\includegraphics[angle=0,width=18.cm]{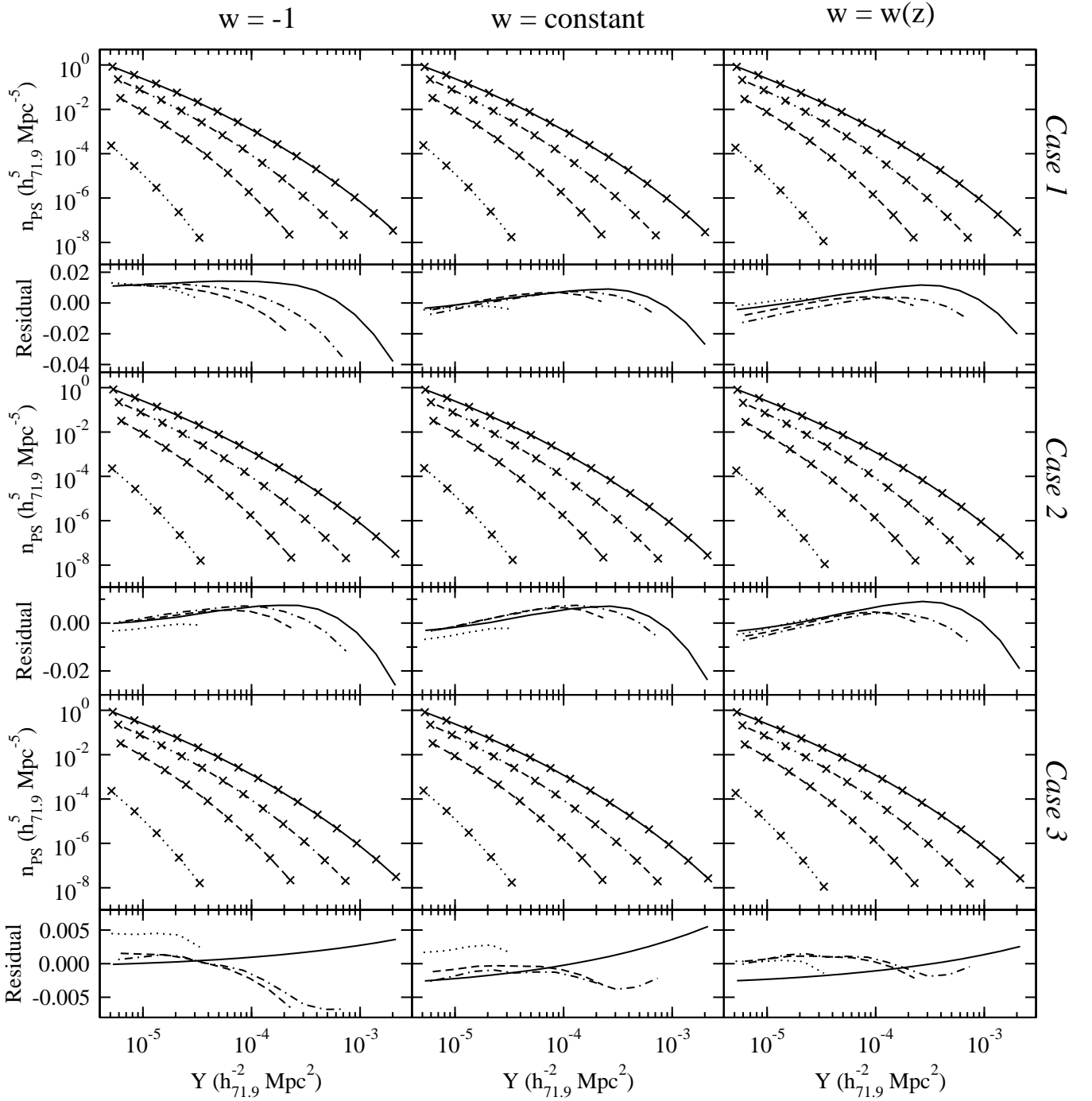}
\caption{
Non-equipartition $Y$ functions (crosses) and the best fitted $Y$ 
functions (lines) at different redshifts are plotted on each figure in 
row 1, 3, and 5 for {\it Case 1}, {\it 2}, and {\it 3}, respectively.  
Residuals in the log of the non-equipartition $Y$ functions are plotted 
under the corresponding figures.  Figures in Column 1 correspond to 
models with dark energy parameters frozen at $w=1$.  Figures in Column 2 
correspond to models with $w$ fitted as constant parameters ($w_1$ frozen 
at zero).  Figures in Column 3 correspond to models with $w$ allowed to 
vary with redshift.  Redshifts at $z=0, 0.5, 1$, and $2$ are shown in 
solid, dash-dotted, dashed, and dotted lines, respectively.
} 
\label{fig:nYfit}
\end{figure*}

Non-equipartition $Y$ functions at several redshifts and 
fitted models are shown in Figure~\ref{fig:nYfit}.  
The deviations in the non-equipartition and 
the fitted $Y$ functions are small and only visible in the residual 
plots.  
In {\it Case 1} and {\it Case 2} with the dark energy equation of state 
parameter frozen at $w=-1$, 
for $z \leq 1$, the non-equipartition $Y$ functions are 
higher than 
the best-fitted $Y$ functions for low-mass clusters, and the opposite is 
true for high-mass clusters.  
At $z=2$, the non-equipartition $Y$ function is slightly higher than the 
best-fitted one in {\it Case 1} with $w$ frozen at $w=-1$, but the 
opposite is true in {\it Case 2}.
The residual is similar in appearance if we free the constant value of 
$w$ ($w_1$ frozen at zero) or allow $w$ to vary with redshift in {\it 
Case 1} 
and {\it Case 2}.  In {\it Case 3}, the residual is more complicated.
The residuals are of the order of a few percent, and this will affect 
the estimated cosmological parameters as discussed below.

The fitted results of the cosmological parameters for different cases we 
considered are summarized in Table~\ref{tab:cosmo}.  In general, for all 
three systematic uncertainty cases we studied, when freezing $w=-1$, 
the deviations of $\Omega_M$ and $\sigma_8$ from the assumed cosmology 
are $\lesssim 1\%$.  The best-fitted parameters happen to be 
consistent with the assumed cosmology, but we can see from the residual 
plots that there are clear systematic deviations by a few percent (e.g., 
Figure~\ref{fig:nYfit}).  Such systematic deviations might be confused 
with the effect of non-Gaussian initial conditions on galaxy cluster mass 
functions, which also show similarly shaped systematic deviations 
\citep[e.g., Figure~1 in][]{FMM09}.

If we free the constant value of $w=w_0$, 
non-equipartition effects can be significant depending on how the 
calibration is done.
For {\it Case 1}, $\Omega_M$, 
$\sigma_8$, and $w_0$ now deviate by $+3.3\%$, $-1.6\%$, and $-5.4\%$, 
respectively.  This shows that ignoring the non-equipartition effects in 
the $Y$ versus mass relation when cross-calibrating
with numerical 
simulations can introduce significant biases in cosmological parameter 
estimations when one is trying to constrain the dark energy equation of 
state.  
Either self-calibrating using a power-law form in the $Y$--$M$ relation 
({\it Case 2}) or calibrating the $Y$--$M$ relation correctly at low 
redshift ({\it Case 3}) can significantly reduce the biases in 
$\Omega_M$, $\sigma_8$, or $w_0$ (down to $\lesssim 1\%$).

If we allow $w$ to vary with redshift, non-equipartition effects can 
introduce significant biases in cosmological parameter estimates (up to 
$\sim 10\%$). 
Self-calibrating using a power-law form in the $Y$--$M$ relation 
({\it Case 2}) can reduce the biases in $\Omega_M$ and $\sigma_8$ down to 
$\lesssim 1\%$, but non-equipartition effects can still introduce a $\sim 
4\%$ bias in the constant normalization of the dark energy equation of 
state parameter ($w_0$) and introduce an apparent evolution of the same 
order.  Calibrating the $Y$--$M$ relation correctly at low redshift ({\it 
Case 3}) can further reduce the bias in $w_0$ to $\lesssim 1\%$, but 
again, there is still an apparent evolution of $\sim 3\%$.
These results show that the mass function technique is very sensitive to 
dark 
energy, and a full understanding of the systematic uncertainties in 
galaxy cluster physics is essential to constrain the dark energy equation 
of state using the mass function technique.

\section{Discussion and Conclusions}
\label{sec:conc}

Numerical simulations have shown that the SZ observables (e.g., the
integrated Comptonization parameter $Y$) for relaxed clusters can be 
biased by a few percent \citep{WS09}, and potentially up to $\sim 
10\%$ in major merging clusters \citep{RN09}.  These results are 
consistent with the SPT and the {\it WMAP} 7 year data which indicate
that electron pressures are smaller than predicted by hydrodynamic 
simulations \citep{Lue+09,Kom+10}.
A few X-ray observations in the 
cluster outer regions also show that the electron pressure is lower than 
the pressure predicted by numerical simulations which assume 
equipartition \citep{Bas+09,GFS+09,Hos+10}.
However, 
it is also 
possible that numerical simulations simply overestimate the gas pressure,
or that the gas is supported in part by other forces such as turbulent or 
cosmic ray pressure.
The $Y$--$M$ relation can be biased if 
non-equipartition effects are not properly taken into account.  Precision 
cosmological studies using the evolution of the galaxy cluster abundance 
rely on a full understanding of the mass--observable relations, 
if the mass of the galaxy clusters cannot be directly measured as is 
usually the case in practice.  We have studied systematically the impact 
of biased $Y$--$M$ relation introduced by the non-equipartition effect 
on SZ surveys and on precision cosmological studies.

While previous studies show that the $Y$--$M$ relation is stable to 
complicated physical processes such as mergers and the power-law index in 
the $Y$--$M$ relation is robust (\citealt{PBM+07}; \citetalias{WSR+08}), 
we have shown that 
non-equipartition effect can introduce a deviation from the power law in 
the $Y$--$M$ relation.  We have fitted a power-law to the 
non-equipartition $Y$--$M$ relation, and we found that there is a $\sim 
1\%$ bias in the power-law index, which is comparable to the uncertainty 
of the power-law index derived from simulations combined with 
observations within $M_{500}$ \citep{APP+09}.  Such a small systematic 
bias and  deviation from the power-law form have important implications for 
SZ surveys and precision cosmological studies using the SZ surveys.

Using the analytic extended Press--Schechter theory to quantify the mass 
function of galaxy clusters, we have studied the non-equipartition effects 
on SZ surveys.  We found that the $Y$ functions can be biased strongly for 
large $Y$ and high-redshift clusters.  For example, the expected number 
of clusters with $Y = (1 \rightarrow 2) \times 10^{-4} 
h_{71.9}^{-2}$~Mpc$^2$ 
[$M = (0.9 \rightarrow 1.3) \times 10^{15}\, h^{-1}_{71.9} M_{\odot}$]
between $z=0.25 \rightarrow 0.75$  can be biased 
by $\sim 5\%$.  The net effect is that ignoring non-equipartition effects 
underestimates the abundance of high-mass and high-redshift clusters.  
Cosmological parameters measured by using the cluster counting technique 
from SZ surveys will be biased if non-equipartition effects are not taken 
into account.  

We have quantified the impact of non-equipartition effects on 
cosmological parameter estimations from SZ surveys by the galaxy cluster 
abundance technique using biased $Y$--$M$ relations.  We considered three 
potential systematic biases in the $Y$--$M$ relations if the 
non-equipartition effect is not properly taken into account during the 
cross-calibration or self-calibration when using the galaxy 
cluster abundance technique.  The best-fit cosmological parameters, 
$\Omega_M,\, \sigma_8$, and also the dark energy equation of state 
parameters [$w=w_0+w_1 z/(1+z)^2$] using the biased $Y$--$M$ relations 
were determined. 
For all the three methods of calibrating the $Y$--mass relation we have 
studied, if the dark energy equation of state parameter is frozen at 
$w=-1$, we find that the best-fit $\Omega_M$ and  $\sigma_8$ are 
consistent with the assumed cosmology to within $\sim 1\%$.  However, 
there are clear systematic deviations of a few percent in the fitted $Y$ 
functions which may be confused with others effects such as  non-Gaussian 
initial conditions \citep{FMM09}.
Models with non-Gaussian initial conditions predict that the actual 
number of clusters with higher mass can be lower than the models with 
Gaussian initial conditions \citep[e.g., Figure~1 in][]{FMM09}; our 
non-equipartition model predicts an apparent smaller $Y$ for high-mass 
clusters and this underestimates the number of high mass clusters if 
the non-equipartition effect is not taken into account.
Note that at such small levels of
systematic deviations, other systematic 
uncertainties such as the use of different mass functions 
\citep[e.g.,][]{PS74, ST99} may introduce larger systematic biases.  
However,  \citet{FMM09} also used the Press--Schechter mass function (as 
we do) to determine the effect of non-Gaussianity.
Thus, we can directly compare these effects with the results of
non-equipartition,
free of biases introduced by the choice of mass function, and
estimate the bias non-equipartition will introduce in efforts to detect
non-Gaussian fluctuations with clusters.

If $w$ is fitted as a constant parameter ($w_1$ frozen at zero), then 
depending 
on the calibration methods, non-equipartition effect can introduce a few 
percent biases on the measured cosmological parameters ({\it Case 1}).  
Either self-calibrating the $Y$--$M$ relation using a power-law form 
({\it Case 2}) or calibrating the $Y$--$M$ relation correctly at low 
redshift ({\it Case 3}) can significant reduce the biases in $\Omega_M$, 
$\sigma_8$, or $w_0$ to $\lesssim 1\%$.  If we allow $w$ to vary 
with redshift, the non-equipartition effect can introduce a bias in 
cosmological parameter of up to $\sim 10\%$ ({\it Case 1}).  In 
particular, 
non-equipartition effects can introduce an apparent evolution in $w$ of a 
few percent in all of the cases we considered.

Using the cluster abundance technique alone, an X-ray survey with 85 
X-ray 
clusters has already constrained some cosmological parameters down to 1\% 
level in statistical uncertainty \citep{Vik+09}.
Ongoing and future SZ surveys will detect thousands of clusters 
\citep[e.g.,][]{Bir99,CHR02,BCM+08}, and this will significantly improve 
the constraints on cosmological parameters.  Therefore, it is important 
to control the systematic uncertainties of galaxy cluster physics at even 
a percentage level.  Hydrodynamic simulations assuming equipartition 
suggest that the integrated $Y$ is a robust mass proxy even when galaxy 
clusters are in the process of merging, and hence the integrated $Y$ is 
taken to be a nearly ideal probe for cosmological studies 
(\citealt{PBM+07}; \citetalias{WSR+08}).

Our results show that if the non-equipartition 
effect 
is not properly taken into account, cosmological parameters can be biased 
significantly (up to $\sim 10\%$).  
In order to take the non-equipartition effect into account when using 
cluster abundance to study precision cosmology, the ultimate solution is 
to include the non-equipartition effect in cosmological simulations 
assuming the non-equipartition physics is known accurately. 
If higher resolution is needed, 
another approach is to correct 
the non-equipartition effect by performing idealized simulations 
\citep[e.g.,][]{WS09} or to re-simulate representative clusters taken from 
cosmological simulations including the non-equipartition effect with 
realistic assumptions.  For the latter case, the non-equipartition effect 
can be taken into account together with other physical processes (e.g., 
gas depletion processes during the formation) which may also affect the $Y$ 
versus mass relation.
However, the above calibration methods by numerical 
simulations rely on the assumption that the non-equipartition physics is 
known accurately, which is in fact not the case at present.  One of the 
key systematic uncertainties is the electron heating efficiency at the 
collisionless shock, $\beta$.  One way to constrain the non-equipartition 
physics is to make direct observations of accretion shocks, which is 
currently not feasible.
We may constrain non-equipartition physics based on observations of 
other astrophysical shocks such as mergers shocks and supernova 
remnants.  However, we have to assume these results apply to cluster 
accretion shocks, which may or may not be the case.  Another route might 
be to perform plasma simulations (e.g., particle-in-cell simulations) to 
constrain the shock physics.  However, to apply the plasma simulation 
results to cluster accretion shocks, a detailed knowledge of the 
pre-shock physics 
such as the magnetic field structure might be needed.  
Clearly, all of the above calculations are necessary to determine the 
range of the systematic uncertainties and the effects of the 
non-equipartition physics, and to constrain the form of the $Y$ versus 
mass relation.  These should be studied in the near future.
Until numerical simulations can directly determine the effects of 
non-equipartition on the  $Y$ versus mass relation from first 
principles, we suggest either to self-calibrate the $Y$ versus 
mass relation using a power-law form at each redshift bin ({\it Case 2}), 
or to calibrate the $Y$ versus mass relation correctly at low redshift 
({\it Case 3}).  These will reduce the biases due to non-equipartition 
on $\Omega_M, \sigma_8$, or 
$w_0$ to better than 1\%.  Important biases introduced by 
other physical processes can be corrected in addition to the 
non-equipartition correction.  However, if one would try to constrain the 
evolution in $w$ to better than 1\%, together with the 
self-calibration method, the constraints from numerical simulations with 
uncertainties less than a percent level might be necessary.

We have shown that using the cluster 
abundance 
to constraint the dark energy equation of state requires a 
full understanding of the systematic uncertainties in galaxy cluster 
physics.   Even though we are only considering the systematic 
uncertainties introduced by the non-equipartition effect, our results 
also suggest that systematic uncertainties in the $Y$--$M$ relation 
introduced by other physics of even a few percent can introduce a 
comparable level of biases in cosmological parameter measurements. 
Future cluster surveys aiming to constrain departures from general relativity
will need to control systematic uncertainties down to a sub-percentage 
level \citep{SVH09}, and hence cluster physics must be understood in a 
comparable accuracy.  
Future theoretical calculations and numerical simulations 
should pay particular attention to the effects of non-thermal physics on 
the electron pressure profiles.  Potential systematic uncertainties 
include conduction, turbulent pressure, magnetic pressure, and 
relativistic pressure supported by cosmic rays.  Deep observations should 
also be carried out to constrain all these effects in detail for 
individual clusters.  The outer regions of galaxy clusters are ideal 
sites for study non-thermal physics.  These studies not only can 
increase our understanding of cosmology, but also can provide information 
on the physics of galaxy clusters and plasma physics under 
extreme conditions.

\acknowledgments
K.W. thanks Avi Loeb and Brian Mason for helpful 
discussions.
Support for this work was provided by the National Aeronautics and Space
Administration, through {\it Chandra} Award Numbers 
TM7-8010X,
GO9-0135X,
and
GO9-0148X,
NASA XMM-Newton Grants
NNX08AZ34G
and
NNX08AW83G,
and NASA Suzaku Grants
NNX08AZ99G,
\linebreak
NNX09AH25G,
and
NNX09AH74G.
We thank the referee for helpful comments.

\clearpage

\begin{deluxetable}{cccccc}
\tabletypesize{\small}
\tablewidth{0pt}
\tablecolumns{6}
\tablecaption{
Effects of Non-Equipartition on the SZ--Mass Relation 
\label{tab:YMfit}
}
\tablehead{
\colhead{} &
\multicolumn{2}{c}{(1--40) $ \times 10^{14} M_{\odot}$} &
&
\multicolumn{2}{c}{(0.1--100) $\times 10^{14} M_{\odot}$} \\
\cline{2-3}
\cline{5-6}
\colhead{$z$} &
\colhead{$\Delta N$} &
\colhead{$\Delta\alpha$} &
&
\colhead{$\Delta N$} &
\colhead{$\Delta\alpha$}
}
\startdata
0.0& 0.977 & -0.0126 &  & 0.967 & -0.0117 \\
0.5& 0.971 & -0.0164 &  & 0.963 & -0.0122 \\
1.0& 0.971 & -0.0152 &  & 0.964 & -0.0109 \\
2.0& 0.970 & -0.0158 &  & 0.969 & -0.0085
\enddata
\end{deluxetable}

\begin{deluxetable}{cccccc}
\tabletypesize{\small}
\tablewidth{0pt}
\tablecolumns{6}
\tablecaption{
Characteristic Numbers of Clusters
\label{tab:Ylimits}
}
\tablehead{
\colhead{$z$} &
\colhead{$z_l$} &
\colhead{$z_u$} &
\colhead{$Y_0 \, (h_{71.9}^{-2}$ Mpc$^2$)} &
\colhead{$M_0 \, (h_{71.9}^{-1} 10^{15} M_{\odot}$)} &
\colhead{$n_{\rm PS,0} \, (h_{71.9}^{5}$ Mpc$^{-5}$)}
}
\startdata
0.0 & 0.00 & 0.25  & $6.10\times 10^{-4}$ & 2.27 & $4.81\times 10^{-6}$\\
0.5 & 0.25 & 0.75  & $5.95\times 10^{-4}$ & 2.24 & $6.01\times 10^{-8}$\\
1.0 & 0.75 & 1.25  & $1.97\times 10^{-4}$ & 1.26 & $4.99\times 10^{-8}$\\
2.0 & 1.25 & 2.75  & $6.04\times 10^{-5}$ & 0.68 & $2.85\times 10^{-10}$
\enddata
\tablecomments{The values of $Y_0$ and $n_{\rm PS,0}$ are defined such that
$\int_{Y_0}^{\infty} \int_{z_l}^{z_u} n_{\rm PS}(Y,z) \, dY \, ( dV / dz ) \, dz = 1$ 
and $n_{\rm PS,0}(z)=n_{\rm PS}(Y_0,z)$ for the equipartition models for different 
redshift intervals $z_l$ to $z_u$.  The value of $M_0$ corresponding to $Y_0$ 
is related by equation~(\ref{eq:YeqvsM}).
}
\end{deluxetable}

\begin{deluxetable}{ccccc}
\tabletypesize{\small}
\tablewidth{0pt}
\tablecolumns{5}
\tablecaption{
Best-Fit Cosmological Parameters and Biases for Different
$Y$ Versus Mass Calibrations
\label{tab:cosmo}
}
\tablehead{
\colhead{Calibration} &
\colhead{$\Omega_M$} &
\colhead{$\sigma_8$} &
\colhead{$w_0$} &
\colhead{$w_1$}
}
\startdata
{\it Case 1} & 0.2548(-1.2\%) & 0.7950(-0.1\%) & [-1]  & [0]\\
& 0.2665(+3.3\%) & 0.7830(-1.6\%) & -0.9464(-5.4\%) & [0]\\
& 0.2680(+3.9\%) & 0.7811(-1.9\%) & -0.9031(-9.7\%) & -0.2362(+5.9\%)\\
\hline
{\it Case 2} & 0.2579(0\%) & 0.7976(+0.2\%) & [-1]  & [0]\\
& 0.2602(+0.9\%) & 0.7951(-0.1\%) & -0.9890(-1.1\%) & [0]\\
& 0.2610(+1.2\%) & 0.7940(-0.3\%) & -0.9577(-4.2\%) & -0.1725(+4.3\%)\\
\hline
{\it Case 3} & 0.2585(+0.2\%) & 0.7953(-0.1\%) & [-1] & [0]\\
& 0.2604(+0.9\%) & 0.7932(-0.4\%) & -0.9910(-0.9\%) & [0]\\
& 0.2601(+0.2\%) & 0.7938(-0.3\%) & -1.0123(+1.2\%) & 0.1167(-2.9\%)
\enddata
\tablecomments{The assumed correct cosmological parameters are 
$\Omega_M=0.258$, $\sigma_8=0.796$, $w_0=-1$, and $w_1=0$.
The bracketed values are the frozen values in the fits. 
The values in parentheses in Columns 2--4 are the percentage 
deviations of the fitted 
cosmological parameters from the assumed parameters.
The values in parentheses in Column 5 are the largest percentage change 
in $w$ between 
the present time ($z=0$) and $z=2$;
this change is $\Delta w = w_1/4$ assuming $w(z)=w_0+w_1 z/(1+z)^2$.
}
\end{deluxetable}


\begin{thebibliography}{}

\bibitem[{{Arnaud} {et~al.}(2009){Arnaud}, {Pratt}, {Piffaretti}, 
{B{\"o}hringer}, {Croston}, \& {Pointecouteau}}]{APP+09}
{Arnaud}, M., {Pratt}, G.~W., {Piffaretti}, R., {B{\"o}hringer}, H., 
{Croston}, J.~H., \& {Pointecouteau}, E. 2009, arXiv:0910.1234

\bibitem[{{Bartlett} {et~al.}(2008) {Bartlett}, {Chamballu}, {Melin}, 
{Arnaud}, \& {Members of the Planck Working Group 5}}]{BCM+08}
{Bartlett}, J.~G., {Chamballu}, A., {Melin}, {J.-B.}, {Arnaud}, M., \& 
{Members of the Planck Working Group 5} 2008, Astron. Nachr., 329, 147

\bibitem[{{Basu} {et~al.}(2009){Basu}}]{Bas+09}
{Basu}, K., et al. 2010, \aap, in press (arXiv:0911.3905)

\bibitem[{{Birkinshaw} (1999){ Birkinshaw}}]{Bir99}
{Birkinshaw}, M. 1999, \physrep, 310, 97

\bibitem[Bond et al.(1991)]{BCE+91} Bond, J.~R., Cole, S., 
Efstathiou, G., \& Kaiser, N.\ 1991, \apj, 379, 440 

\bibitem[{{Carlstrom} {et~al.}(2002){Carlstrom}, {Holder}, \&
{Reese}}]{CHR02}
{Carlstrom}, J.~E., {Holder}, G.~P., \& {Reese}, E.~D. 2002, \araa, 40,
643

\bibitem[{{Ettori \& Fabian}(1998){Ettori}, \& {Fabian}}]{EF98}
{Ettori}, S., \& {Fabian}, A.~C. 1998, \mnras, 293, L33


\bibitem[{{Fedeli} {et~al.}(2009){Fedeli}, {Moscardini}, \& 
{Matarrese}}]{FMM09}
{Fedeli}, C., {Moscardini}, L., \& {Matarrese}, S.
2009, \mnras, 397, 1125

\bibitem[{{Fox \& Loeb}(1997){Fox}, \& {Loeb}}]{FL97}
{Fox}, D.~C., \& {Loeb}, A. 1997, \apj, 491, 459

\bibitem[{{George} {et~al.}(2009){George}, {Fabian}, {Sanders}, {Young},
\& {Russell}}]{GFS+09}
{George}, M.~R., {Fabian}, A.~C., {Sanders}, J.~S., {Young}, A.~J., \&
{Russell}, H.~R. 2009, \mnras, 395, 657


\bibitem[{{Haiman} {et~al.}(2001){Haiman}, {Mohr}, \& {Holder}}]{HMH01}
{Haiman}, Z., {Mohr}, J.~J., \& {Holder}, G.~P. 2001, \apj, 553, 545

\bibitem[{{Holder} {et~al.}(2007){Holder}, {McCarthy}, \& 
{Babul}}]{HMB07} 
{Holder}, G.~P., {McCarthy}, I.~G., \& {Babul}, A.\ 2007, \mnras, 382, 
1697

\bibitem[{{Hoshino} {et~al.}(2010){Hoshino}}]{Hos+10}
{Hoshino}, A., et al. 2010, \pasj, 62, 371

\bibitem[{{Hu}(2003)}]{Hu03}
{Hu}, W. 2003, \prd, 67, 081304

\bibitem[{{Komatsu} {et~al.}(2010){Komatsu}}]{Kom+10}
{Komatsu}, E., et al. 2010, arXiv:1001.4538

\bibitem[{{Lacey} \& {Cole}(1993)}]{LC93}
{Lacey}, C., \& {Cole}, S. 1993, \mnras, 262, 627

\bibitem[{{Levine} {et al.}(2002){Levine}, {Schulz}, \& {White}}]{LSW02}
{Levine}, E.~S., {Schulz}, A.~E., \& {White}, M.
2002, \apj, 577, 569

\bibitem[{{Loeb} (2002){Loeb}}]{Loe02}
{Loeb}, A. 2002, New Astronomy, 7, 279


\bibitem[{{Lueker} {et~al.}(2009){Lueker}}]{Lue+09}
{Lueker}, M., et al. 2009, arXiv:0912.4317

\bibitem[{{Luki{\'c}} {et al.}(2007){Luki{\'c}}, {Heitmann}, {Habib}, 
{Bashinsky}, \& {Ricker}}]{LHH+07} 
Luki{\'c}, Z., Heitmann, K., Habib, S., Bashinsky, S., \& Ricker, P.~M.\ 
2007, \apj, 671, 1160

\bibitem[{{Majumdar} \& {Mohr}(2003)}]{MM03}
{Majumdar}, S., \& {Mohr}, J.~J.
2003, \apj, 585, 603

\bibitem[{{Mantz} {et~al.}(2008){Mantz}, {Allen}, {Ebeling}, \& 
{Rapetti}}]{MAE+08}
{Mantz}, A., {Allen}, S.~W., {Ebeling}, H., \& {Rapetti}, D.
2008, \mnras, 387, 1179

\bibitem[{{Melin} {et~al.}(2005){Melin}, {Bartlett}, \& 
{Delabrouille}}]{MBD05}
{Melin}, {J.-B.}, {Bartlett}, J.~G., \& {Delabrouille}, J.
2005, \aap, 429, 417

\bibitem[{{Poole} {et~al.}(2007) {Poole}, {Babul}, {McCarthy}, {Fardal}, 
{Bildfell}, {Quinn}, \& {Mahdavi}}]{PBM+07}
{Poole}, G.~B., {Babul}, A., {McCarthy}, I.~G., {Fardal}, M.~A., 
{Bildfell}, C.~J., {Quinn}, T, \& {Mahdavi}, A. 2007, \mnras, 380, 437

\bibitem[{{Press} \& {Schechter}(1974)}]{PS74}
{Press}, W.~H., \& {Schechter}, P. 1974, \apj, 187, 425

\bibitem[{{Randall} {et~al.}(2002){Randall}, {Sarazin}, 
\& {Ricker}}]{RSR02}
{Randall}, S.~W., {Sarazin}, C.~L., \& {Ricker}, P.~M. 2002, \apj, 577, 
579 (RSR)

\bibitem[{{Reid \& Spergel}(2006){Reid}, \& {Spergel}}]{RS06}
{Reid}, B.~A., \& {Spergel}, D.~N. 2006, \apj, 651, 643

\bibitem[{{Rudd \& Nagai}(2009){Rudd}, \& {Nagai}}]{RN09}
{Rudd}, D.~H., \& {Nagai}, D. 2009, \apjl, 701, L16

\bibitem[{{Schmidt} {et~al.}(2009) {Schmidt}, {Vikhlinin},\& 
{Hu}}]{SVH09}
{Schmidt}, F., {Vikhlinin}, A., \& {Hu}, W. 2009, \prd, 80, 083505

\bibitem[Sheth \& Tormen(1999)]{ST99} Sheth, R.~K., \& 
Tormen, G.\ 1999, \mnras, 308, 119

\bibitem[{{Springel} {et~al.}(2005){Springel}, {White}, {Jenkins}, 
{Frenk}, {Yoshida}, {Gao}, {Navarro}, {Thacker}, {Croton}, {Helly}, 
{Peacock}, {Cole}, {Thomas}, {Couchman}, {Evrard}, {Colberg}, \& 
{Pearce}}]{Spr+05}
Springel, V., et al.\ 
2005, \nat, 435, 629



\bibitem[{{Vikhlinin} {et~al.}(2009){Vikhlinin}, {Burenin}, {Ebeling}, 
{Forman}, {Hornstrup}, {Jones}, {Kravtsov}, {Murray}, {Nagai}, 
{Quintana}, \& {Voevodkin}}]{Vik+09}
{Vikhlinin}, A., et al.\
2009, \apj, 692, 1033

\bibitem[{{Wik} {et~al.}(2008){Wik}, {Sarazin}, {Ricker}, \& 
{Randall}}]{WSR+08}
{Wik}, D.~R., {Sarazin}, C.~L., {Ricker}, P.~M., \& {Randall}, 
S.~W. 2008, \apj, 680, 17 (WSR)

\bibitem[{{Wong \& Sarazin}(2009){Wong} \& {Sarazin}}]{WS09}
{Wong}, {K.-W.}, \& {Sarazin}, C.~L. 2009, \apj, 707, 1141


\end{thebibliography}
\end{document}